\documentclass[aps,prc,preprintnumbers,twocolumn,superscriptaddress,floatfix,showpacs]{revtex4}

\usepackage{graphicx}

\begin{document}


\title{Characteristics of the fragments produced in central collisions of
$^{129}$Xe+$^{nat}$Sn from 32 to 50 AMeV}

\author{S.~Hudan}
\altaffiliation[Present address: ]{Department of Chemistry, Indiana University, Bloomington, IN
47405, USA}
\affiliation{ GANIL, CEA et IN2P3-CNRS, B.P.~5027, F-14076 Caen Cedex, France.}
\author{A.~Chbihi}
\affiliation{ GANIL, CEA et IN2P3-CNRS, B.P.~5027, F-14076 Caen Cedex, France.}
\author{J.D.~Frankland}
\affiliation{ GANIL, CEA et IN2P3-CNRS, B.P.~5027, F-14076 Caen Cedex, France.}
\author{A.~Mignon}
\affiliation{ GANIL, CEA et IN2P3-CNRS, B.P.~5027, F-14076 Caen Cedex, France.}
\author{J.P.~Wieleczko}
\affiliation{ GANIL, CEA et IN2P3-CNRS, B.P.~5027, F-14076 Caen Cedex, France.}
\author{G.~Auger}
\affiliation{ GANIL, CEA et IN2P3-CNRS, B.P.~5027, F-14076 Caen Cedex, France.}
\author{N.~Bellaize}
\affiliation{ LPC, IN2P3-CNRS, ISMRA et Universit\'e, F-14050 Caen Cedex, France.}
\author{B.~Borderie}
\affiliation{ Institut de Physique Nucl\'eaire, IN2P3-CNRS, F-91406 
Orsay Cedex,  France.}
\author{R.~Bougault}
\affiliation{ LPC, IN2P3-CNRS, ISMRA et Universit\'e, F-14050 Caen Cedex, France.}
\author{B.~Bouriquet}
\affiliation{ GANIL, CEA et IN2P3-CNRS, B.P.~5027, F-14076 Caen Cedex, France.}
\author{A.M.~Buta}
\affiliation{ LPC, IN2P3-CNRS, ISMRA et Universit\'e, F-14050 Caen Cedex, France.}
\author{J.~Colin}
\affiliation{ LPC, IN2P3-CNRS, ISMRA et Universit\'e, F-14050 Caen Cedex, France.}
\author{D.~Cussol}
\affiliation{ LPC, IN2P3-CNRS, ISMRA et Universit\'e, F-14050 Caen Cedex, France.}
\author{R.~Dayras}
\affiliation{ DAPNIA/SPhN, CEA/Saclay, F-91191 Gif sur Yvette, France.}
\author{D.~Durand}
\affiliation{ LPC, IN2P3-CNRS, ISMRA et Universit\'e , F-14050 Caen Cedex, France.}
\author{E.~Galichet}
\affiliation{ Institut de Physique Nucl\'eaire, IN2P3-CNRS, F-91406 Orsay Cedex,
 France.}
\affiliation{ Conservatoire National des Arts et M\'etiers, F-75141 Paris
Cedex 03.}
\author{D.~Guinet}
\affiliation{ Institut de Physique Nucl\'eaire, IN2P3-CNRS et Universit\'e
F-69622 Villeurbanne, France.}
\author{B.~Guiot}
\affiliation{ GANIL, CEA et IN2P3-CNRS, B.P.~5027, F-14076 Caen Cedex, France.}
\author{G.~Lanzalone}
\altaffiliation[Permanent address: ]{Laboratorio Nazionale del Sud, 
Via S. Sofia 44, I-95123 Catania, Italy.}
\affiliation{ Institut de Physique Nucl\'eaire, IN2P3-CNRS, F-91406 Orsay Cedex,
 France.}
\author{P.~Lautesse}
\affiliation{ Institut de Physique Nucl\'eaire, IN2P3-CNRS et Universit\'e
F-69622 Villeurbanne, France.}
\author{F.~Lavaud}
\altaffiliation[Present address: ]{ DAPNIA/SPhN, CEA/Saclay, F-91191 Gif sur Yvette, France.}
\affiliation{ Institut de Physique Nucl\'eaire, IN2P3-CNRS, F-91406 Orsay Cedex,
 France.}
\author{J.F.~Lecolley}
\affiliation{ LPC, IN2P3-CNRS, ISMRA et Universit\'e, F-14050 Caen Cedex, France.}
\author{R.~Legrain}
\thanks{deceased}
\affiliation{ DAPNIA/SPhN, CEA/Saclay, F-91191 Gif sur Yvette, France.}
\author{N.~Le~Neindre}
\affiliation{ GANIL, CEA et IN2P3-CNRS, B.P.~5027, F-14076 Caen Cedex, France.}
\author{O.~Lopez}
\affiliation{ LPC, IN2P3-CNRS, ISMRA et Universit\'e, F-14050 Caen Cedex, France.}
\author{L.~Manduci}
\affiliation{ LPC, IN2P3-CNRS, ISMRA et Universit\'e, F-14050 Caen Cedex, France.}
\author{J.~Marie}
\affiliation{ LPC, IN2P3-CNRS, ISMRA et Universit\'e, F-14050 Caen Cedex, France.}
\author{L.~Nalpas}
\affiliation{ DAPNIA/SPhN, CEA/Saclay, F-91191 Gif sur Yvette, France.}
\author{J.~Normand}
\affiliation{ LPC, IN2P3-CNRS, ISMRA et Universit\'e, F-14050 Caen Cedex, France.}
\author{M.~P\^arlog}
\affiliation{ National Institute for Physics and Nuclear Engineering, RO-76900
Bucharest-M\u{a}gurele, Romania.}
\author{P.~Paw{\l}owski}
\affiliation{ Institut de Physique Nucl\'eaire, IN2P3-CNRS, F-91406 Orsay Cedex,
 France.}
\author{M.~Pichon}
\affiliation{ LPC, IN2P3-CNRS, ISMRA et Universit\'e, F-14050 Caen Cedex, France.}
\author{E.~Plagnol}
\affiliation{ Institut de Physique Nucl\'eaire, IN2P3-CNRS, F-91406 Orsay Cedex,
 France.}
\author{M.~F.~Rivet} 
\affiliation{ Institut de Physique Nucl\'eaire, IN2P3-CNRS, F-91406 
Orsay Cedex,  France.} 
\author{E.~Rosato}
\affiliation{ Dipartimento di Scienze Fisiche e Sezione INFN, Univ. di 
Napoli ``Federico II'', I-80126 Napoli, Italy.}
\author{R.~Roy}
\affiliation{ Laboratoire de Physique Nucl\'eaire, Universit\'e Laval,
Qu\'ebec, Canada.}
\author{J.C.~Steckmeyer}
\affiliation{ LPC, IN2P3-CNRS, ISMRA et Universit\'e, F-14050 Caen Cedex, France.}
\author{G.~T\u{a}b\u{a}caru} 
\affiliation{ National Institute for Physics and Nuclear Engineering, RO-76900
Bucharest-M\u{a}gurele, Romania.}
\author{B.~Tamain}
\affiliation{ LPC, IN2P3-CNRS, ISMRA et Universit\'e, F-14050 Caen Cedex, France.}
\author{A.~van Lauwe}
\affiliation{ LPC, IN2P3-CNRS, ISMRA et Universit\'e, F-14050 Caen Cedex, France.}
\author{E.~Vient}
\affiliation{ LPC, IN2P3-CNRS, ISMRA et Universit\'e, F-14050 Caen Cedex, France.}
\author{M.~Vigilante}
\affiliation{ Dipartimento di Scienze Fisiche e Sezione INFN, Univ. di 
Napoli ``Federico II'', I-80126 Napoli, Italy.}
\author{C.~Volant}
\affiliation{ DAPNIA/SPhN, CEA/Saclay, F-91191 Gif sur Yvette, France.}

\collaboration{INDRA collaboration} \noaffiliation

\date{16 October 2002}
\begin{abstract}
Characteristics of the primary fragments produced in central 
collisions of $^{129}$Xe + $^{nat}$Sn from 32 to 50 AMeV have been obtained.
By using the 
correlation technique for the relative velocity between light charged 
particles (LCP) and fragments, we were able to extract the 
multiplicities and average kinetic energy of secondary evaporated LCP. 
We then reconstructed the size and excitation energy of the primary fragments. 
For each bombarding energy a constant value of the excitation 
energy per nucleon over the whole range of fragment charge has been found.
This value saturates at 3 AMeV for beam energies 39 AMeV and above.
The corresponding secondary evaporated LCP represent less than 40\%
of all produced particles and decreases down to 23\% for 50 AMeV. 
The experimental characteristics of the primary fragments are compared to the predictions
of statistical multifragmentation model (SMM) calculations. 
Reasonable agreement between the data and the calculation has been found for any
given incident energy. However SMM fails to reproduce the trend of the
excitation function of the primary fragment excitation energy and the 
amount of secondary evaporated LCP's. 
\end{abstract}


\maketitle

\section{Introduction}
Multiple intermediate mass fragment (IMF) production in central heavy-ion collisions
is related to the properties of nuclear matter under
extreme conditions. Many different models have been proposed in order to explain
the observed fragment production, and both theoretically and experimentally the situation 
is not clear. Models with widely differing basic hypotheses can be equally good at describing
the same data such as charge distributions, mean energies and angular
distributions. In order to gain further understanding it is therefore
necessary to have more detailed information on the multifragmentation process.

One aspect of the reactions for which different models give very different
predictions are the excitation energies of what we will call from now on the
`primary fragments': in other words the nuclei present around $\sim$100 fm/c
after the collision, which are not necessarily the same as those arriving in
the detectors a few 10's of ns later. In Quantum Molecular Dynamics
~\cite{QMD:main-ref,Neb99:qmd-xesn-radialflow,I16-Neb99,Tirel:PhD}
simulations or Microcanonical Metropolis Monte-Carlo model ~\cite{MMMC:main-ref}
calculations the primary fragments are rather cold i.e. they are almost unaffected by
subsequent secondary decays and arrive unchanged in the detectors.
In the former case, the (lack of) excitation energy in the nascent fragments is determined
by the collision dynamics, whereas in the latter case it is an assumption of the
model when calculating the statistical weights of the partitions.
On the other hand, Antisymmetrised Molecular Dynamics 
~\cite{AMD:main-ref1,AMD:main-ref2,AMD:main-ref3} and
Stochastic Mean Field ~\cite{BOB:main-ref1,BOB:main-ref2}
simulations both predict moderately ``hot'' primary
fragments in reactions around the Fermi energy, with $E^*_{pr}\sim 2-3$ AMeV
~\cite{Hudan:PhD,Hud02:comp-amd-data,I29-Fra01}. Finally, the Statistical
Multifragmentation Model (SMM)~\cite{SMM:main-ref} and the microcanonical
multifragmentation model of ~\cite{Raduta:micro1,Raduta:micro2}
allow primary fragments to be excited, the actual value in any given calculation
being determined by energy conservation
and the statistical weight given by the
associated level density parametrisation. This latter may or may not take into
account the level density limitation in isolated nuclei at high excitation
~\cite{Dean:high_e_levdens}, equivalent to excluding from the primary partitions
levels with very short lifetimes or introducing an effective limiting
(maximum) temperature for hot nuclei ~\cite{Levit:T_limite,Koonin:T_limite}.

Our previous experimental work~\cite{I11-Mar98} has shown that the 
 reconstruction of the average size and excitation energy of the
primary fragments is possible by means of
fragment-light charged particles (IMF-LCP) relative velocity correlation functions. A constant
value of the excitation energy of the primary fragments has been
deduced at about 3 AMeV for the Xe + Sn system at 50 AMeV. It was also 
possible to deduce the multiplicities of the secondary particles evaporated 
by the primary fragments. More recently analogous results and conclusions have
 been obtained for central collisions of Kr + Nb at 45 AMeV~\cite{Staszel:2001rw}.  
An important question arises from these studies :  
what is the evolution of the fragment excitation energies and secondary LCP
multiplicities as a function of incident energy ? 
The experimental answer to this question may permit to distinguish between different
scenarios and assumptions made by different models.
It should give a strong test of the validity of some of their basic hypotheses. 

In this paper we extend the previous study~\cite{I11-Mar98} to a wider 
incident energy range, from 32 to 50 AMeV for central collisions of 
the Xe + Sn system measured with the 4$\pi $ INDRA 
detector~\cite{I3-Pou95,I4-Ste95,I5-Pou95}. 
Excitation functions for the fragment excitation energy and 
the fraction of secondary emitted LCP correlated to the fragments will be
shown. We will give in section 2 a brief description of the detector, the way 
we select the events and an overview of the fragment production.
We will describe in section 3 the method employed to 
extract the LCP's correlated to each fragment. The method used for the
decorrelation in this work is different from the previous one~\cite{I11-Mar98} but
gives almost the same results.
The experimental results are then given in section 4. 
In section 5 a comparison of the deduced primary excitation energy and secondary 
LCP multiplicities to SMM calculations is given. We then discuss the results
in section 6.

\section{Experiment}

\subsection{Experimental set-up}

The experiment was performed at GANIL with the multidetector INDRA ~\cite
{I3-Pou95,I4-Ste95,I5-Pou95}. This charged product detector covers about 90\% of the 4$\pi$
solid angle. The total number of detection cells is 336 arranged according
to 17 rings centred on the beam axis. The first ring ($2^o$-$3^o$) is made
of fast NE102/NE115 phoswich detectors. Rings 2 to 9 cover the angular
range from $3^o$ to $45^o$ and are made of three detector layers : a low
pressure gas-ionisation chamber, a 300 $\mu m$ thick silicon detector and a
14 to 10 cm thick CsI(Tl) scintillator. The remaining 8 rings cover the
angular range from $45^o$ to $176^o$ and have two detection layers :
ionisation chamber and 7.6 to 5 cm thick CsI(Tl) scintillators. For the
studied system Xe + Sn, fragments with Z up to 54 are identified in the
forward region. Beyond $45^o$, the charge resolution is one unit up to
Z=16 and few charges above. Over the whole angular range, a very good isotope 
identification is obtained for Z=1 to Z=3, except for particles with low 
energies where ambiguities are unresolved.

The energy calibration of the CsI(Tl) scintillators was obtained for light
charged particles (LCP) by means of the elastic and inelastic scattering of
secondary LCP beams ($p,d,t$,$^{3}He$,$^{4}He$) produced by the
fragmentation of a 95 AMeV $^{16}$O beam on a thick C target. These
particles were then momentum selected by the ``alpha magnetic spectrometer''
of GANIL and scattered in a C or Ta target installed in the INDRA reaction chamber.
For $Z\geq 3$ fragments, the energy calibration was made by using the $%
\Delta E$/E technique. A typical energy resolution was about $4\%$. The
energy threshold was a few 100 keV for light particles, 0.7 AMeV for Z=3 and
1.4 AMeV for Z=35. A complete technical description of INDRA, its
calibration and its electronics can be found in ~\cite{I3-Pou95,I4-Ste95,I5-Pou95,
I14-Tab99,I33-Par02,I34-Par02}.

\subsection{Selection of central collisions}

Two selections have been made to isolate central collisions. The first
one is the requirement of quasi-complete events by accepting in the off-line
analysis only
events having total detected charge $(Z_{tot})$ $\geq $ $80\%$ of the
initial total charge of the system. The second is the use of the 
flow angle ($\theta _{flow}$) 
selection~\cite{Lec94:theta_flow}. 
This angle is a global observable defined as the
angle between the beam axis and the main direction of emission of matter in
each event as determined by the energy tensor calculated from fragment ($Z\geq 3$)
c.m. momenta~\cite{Cugnon:1983gv}. It has been shown for heavy ion
reactions in the Fermi energy range
\cite{Lec94:theta_flow,I9-Mar97,I28-Fra01} that events with small $\theta _{flow}$ are
dominated by binary dissipative collisions. On the other hand events with
little or no memory of the entrance channel should be isotropic, thus favouring
large $\theta _{flow}$ ($P(\theta_{flow})\sim \sin\theta_{flow}$). Quasi-complete
events having $\theta_{flow}\geq 45^\circ$ for 50 AMeV bombarding energy and
$\theta_{flow}\geq 60^\circ$ for the three other systems
correspond to an isotropic emission of the IMF in the centre of mass 
of the whole system. These events are compatible with decay of a compact object which
could take place after fast emission of a direct light particle component.
Indeed the velocity of the fragments are evenly distributed around the centre 
of mass velocity~\cite{Salou:PhD}. By taking into account the detection
efficiency and other biases due to the selection we have estimated the cross
sections for `isotropic central collisions' to decrease from 115$\pm$20 mb at
32 AMeV to 85$\pm$10 mb at 50 AMeV. More details about this
event selection for Xe + Sn collisions at 32--50 AMeV incident energy and the
extraction of the cross sections can be found in~\cite{Salou:PhD}.

\subsection{Overview of fragment production in central collisions}

\begin{figure}
\includegraphics[width=\columnwidth]{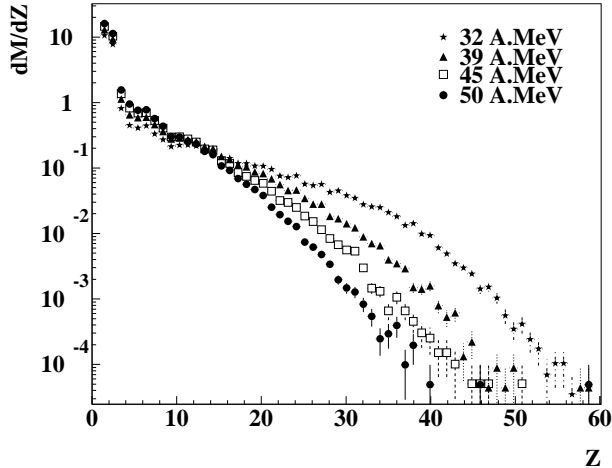}%
\caption{Charge distributions of fragments produced in central collisions 
of Xe + Sn at 4 incident energies: 32, 39, 45 and 50 AMeV.
\label{f:zdis}}
\end{figure}

\begin{figure}
\includegraphics[width=\columnwidth]{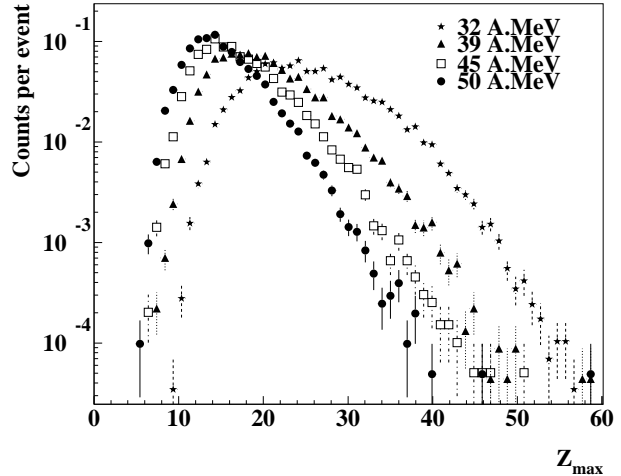}%
\caption{Charge distributions of the heaviest fragment per event produced 
in central collisions of Xe + Sn at 4 incident energies: 32, 39, 45 
and 50 AMeV.
\label{f:zmax1}}
\end{figure}

Before determining the characteristics of the fragments, let us first show 
an overview of their production in central collisions of Xe +
Sn from 32 to 50 AMeV. Figure \ref{f:zdis} shows their charge distributions
normalised to the number of events so that the four bombarding energies can
be compared. The production of small fragments ($Z\le 10$) increases with
incident energy.
For the charge range 10 to 15 the four distributions exhibit a kind of 
``plateau''. In this range the fragment production rates are roughly equivalent
 whatever the incident
energy is. Finally the charge distributions evolve from a broad shape at lower
incident energy, where residues up to the size of the projectile are
observed, toward an almost exponential form at 50 AMeV, favouring the production
of lighter fragments. Moreover figure \ref{f:zmax1}, 
where the distributions
of the heaviest fragment in the event are shown, confirms this behaviour. Here
again the distribution at 32 AMeV is very broad, its average value is
$<Z_{max}> = 25$, it decreases to smaller $<Z_{max}> = 15$ 
at 50 AMeV. It is important to notice
that, even with this strong evolution in the charge distribution, the mean fragment 
multiplicity does not change too much with the incident energy. It evolves from
5 to 7 fragments with $Z\geq 3$ only. 

\begin{figure}
\includegraphics[width=\columnwidth]{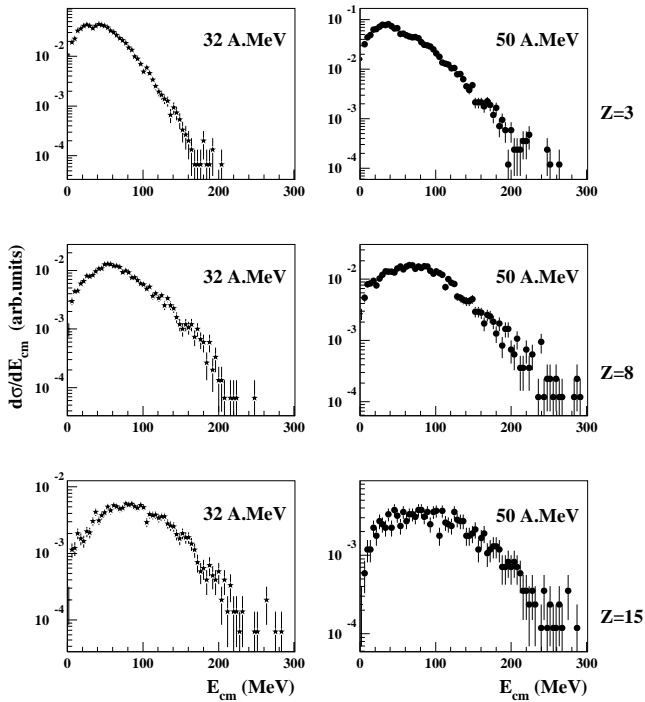}%
\caption{Angle-integrated centre of mass kinetic energy spectra of Li, O and P produced 
in central collisions of Xe + Sn at 32 and 50 AMeV.
\label{f:spectre_ini}}
\end{figure}

Concerning the kinematic characteristics of the fragments, figure 
\ref{f:spectre_ini} shows an example of the fragment angle-integrated
centre of mass kinetic       
energy spectra for Li, O and P nuclei produced in central collisions of Xe + Sn
at 32 and 50 AMeV. The distributions are broad; they are broader for the
heavier elements. Comparing the 
spectra obtained at 32 and 50 AMeV, we observe easily that their shape, 
in particular
the slopes of their exponential tails, are different. The distributions are broader and
harder at 50 than at 32 AMeV.

\begin{figure}
\includegraphics[width=\columnwidth]{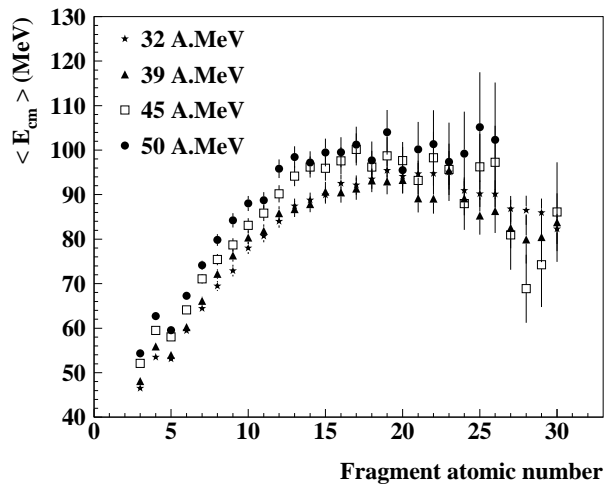}%
\caption{Mean centre of mass kinetic energy of the fragments produced 
in central collisions ($\theta_{flow}$ selection, see text)
of Xe + Sn at 4 incident energies: 32, 39, 45 
and 50 AMeV as a function of their atomic number. The statistical error bars are shown.
\label{f:ecmallz2}}
\end{figure}

We finally present in figure \ref{f:ecmallz2}, for the four incident energies, the 
mean centre of mass kinetic energy of the fragments as a function of their atomic number. It
increases with the charge $Z$ and then saturates beyond $Z = 15$. It also
increases with the bombarding energy but very little. We wondered whether this
observation is true for central collisions in general, or is rather dependent on our selection.
In fact it is the mean kinetic 
energy of the heaviest fragment which ``saturates'' while that of the other fragments increases
monotonously with $Z$. The $\theta_{flow}$ selection we use is derived from fragment kinetic properties
and therefore its effect on observables such as e.g. fragment energies and angular
distributions must be taken into account in events selected in this way.  
Nevertheless this selection has little influence on the study of individual
fragment characteristics such as excitation energy and secondary decay,
whatever the mechanism of their formation.   

\section{Extraction of secondary evaporated light charged particles
\label{s:extraction}}

The main aim of this work is to extract the intrinsic properties of the fragments
independently of the mechanism responsible for their formation. Are they excited~?
If so what are the associated LCP evaporated from the parents~? 
Reconstructing the primary fragments
assumes that we are able experimentally to isolate the secondary contribution.
This is possible if the fragments formed are not too excited, 
so that the time scale associated with their decay is much greater than the 
time scale of their production. The origin of the fragments is still an open question 
but is not the subject of this paper.

\subsection{Correlation functions}

In the previous section it was shown that on average about
6 fragments are produced in central collisions of Xe + Sn at different energies. 
However the
production of LCP is much more important, on average their number reaches
28 particles for the 50 AMeV beam energy. There are at least three different
stages to produce these particles : i) in the early stage of the collision, in
this case we call them primary particles; ii) at the same time as the
formation of the fragments; iii) they can be emitted from the excited primary
fragments, we call those the secondary particles.  
Correlation functions are a powerful tool for extracting small signals. 
This is the method we used to
extract, on the average, the LCP emitted from each 
fragment. With the help of simulations we have developed a correlation 
technique to extract possible signals~\cite{I11-Mar98,hudan:Bormio00,Hudan:PhD}. 

\begin{figure}
\includegraphics[width=\columnwidth]{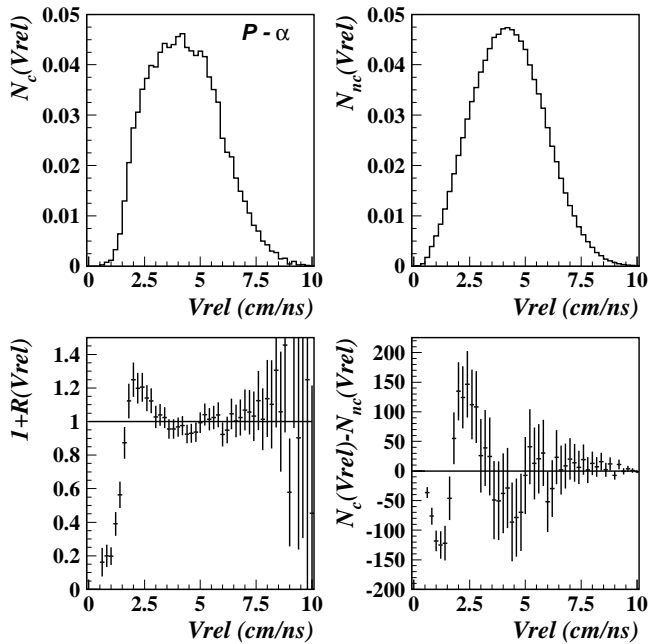}%
\caption{Relative velocity spectra of P-$\alpha$ pairs observed for the Xe+Sn
system at 32 AMeV. Top-left hand panel corresponds to the correlated events, top-right
hand panel the uncorrelated events (mixed events), bottom-left hand panel the
correlation function and bottom-right panel shows the difference function.
\label{f:data}}
\end{figure}

Figure \ref{f:data} shows the relative velocity distributions : 
i) for P-$\alpha$ pairs taken from the same events, 
ii) for the uncorrelated events obtained by taking 
the fragment from a given event and the light particle from another event, iii)
the correlation function defined as the ratio of the correlated and 
uncorrelated relative velocity distributions, iv) the difference correlation function
defined as the difference between the correlated and uncorrelated distributions. 
In this work to decorrelate the relative velocity between the fragment and the LCP
pairs we used the event mixing 
procedure~\cite{Drijard:1984pe}. In this example,
for each phosphorus found in an event having a number of 
alphas $N_a$
we take randomly $N_a$ alphas emitted in $N_a$ other events. 

This technique is
different from the one reported in ref.~\cite{I11-Mar98} where Li nuclei
were used to decorrelate the events. The problem with such a technique is that
the Li can be the product of the known resonance of $^7Be$ which decays to $^6Li$
+ $p$ and increases the background, thus decreasing the yield of true correlated
protons.  
However the final result is almost the same (within the error bars)
as the old method of decorrelation of events based on Li.
  
As we can see, the example presented in figure \ref{f:data}, exhibits a bump
around 2.5 cm/ns relative velocity in the correlation function and difference function
which may be related to
the evaporation of an $\alpha$ particle from a parent of phosphorus. 
 The behaviour of this correlation encourages us to make such an
analysis. However it is necessary to simulate the background in order to extract
the signal.  

\subsection{Simulation of the background shape}\label{p:sim_back}

The objective of this simulation is not to reproduce the data, it is more
to have an idea about the shape of the background. 
We used a modified version of the SIMON event generator ~\cite{simon-durand} to 
simulate a 
scenario deduced from BNV ~\cite{BNV:main-ref} calculations. Two steps are assumed in these
simulations. The first step is the cooling of the initial fused system through
a sequential light particles (LP) emission process (primary LP),
the second is the fragmentation of the smaller remaining source where the
remaining excitation energy is shared between a fixed number of primary 
fragments (typically 6 to 7 fragments). Then the primary fragments decay 
sequentially while moving apart under Coulomb forces plus an initial
radial velocity. This simulation reproduces reasonably well the global 
experimental features. In particular the kinematic observables are well
reproduced (see for example ref.~\cite{I9-Mar97}.) 
\begin{figure}
\includegraphics[width=\columnwidth]{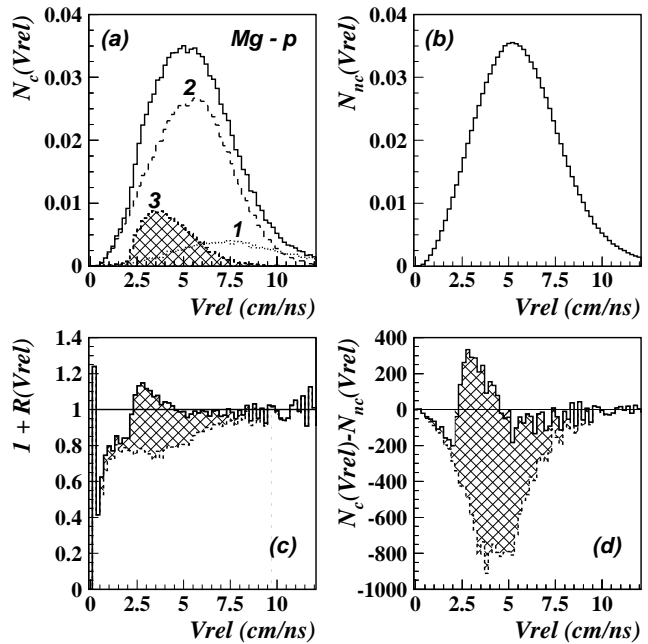}%
\caption{Relative velocity spectra of Mg-p pairs simulated for the Xe+Sn system
at 32 AMeV. (a)For correlated events: total spectrum (thick histogram), 
contribution of primary emission (dotted histograms, high energy
contribution), secondary emission from primary fragments which produce Mg nuclei
(hatched-dashed) and do not produce the considered fragment (dashed histograms). 
(b)For uncorrelated events. 
(c) The correlation function (continuous histogram), the real 
background (dashed histogram) and the contribution from the secondary 
emission from the parents of the Mg fragments (hatched area) are shown. 
(d) The difference function is shown here. The notations are similar 
to (c).
\label{f:simon}}
\end{figure}

The calculated relative velocities are shown in Figure \ref{f:simon}.a (thick lines)
for Mg-p pairs and for input parameters which reproduce data for 
the 32 AMeV Xe + Sn central collisions. 
Since in this version of SIMON we know which particle is emitted from which fragment, 
we plotted in the same figure the
different contributions : the primary contribution (dotted histogram) that we call contribution 1, 
the evaporated protons from all other
fragments except the parents of magnesium (dashed histogram) that we call contribution 2 
and finally the protons 
emitted from the parents of
detected magnesium fragments (hatched-dashed histogram) called contribution 3. As expected, the latter 
contribution is very small, it represents the protons truly correlated 
to the magnesium nucleus that we must extract from the data. Figure \ref{f:simon}.b
 shows the 
uncorrelated relative velocity for Mg-p pairs reconstructed by mixing 
the calculated events.    
Figure \ref{f:simon}.c and \ref{f:simon}.d show the Mg-p correlation
function (the ratio of the correlated and uncorrelated relative velocity distributions
 of fig.\ref{f:simon}.a and fig.\ref{f:simon}.b) 
and the difference function (the difference of the latter distributions), respectively. 
In the same figures are plotted the 
associated true backgrounds (dashed histograms) calculated by dividing (subtracting) the sum of
contributions 1 and 2 by the uncorrelated distribution (of fig.\ref{f:simon}.b).  
The hatched areas represent 
the contribution of secondary emission from the parents of magnesium (contribution 3). The 
shape of the background shown in figure \ref{f:simon}.c is well 
fitted by the function :
\begin{equation}
 R(V_{rel}) = A - {1 \over B V_{rel} + C}
\end{equation}
where A, B and C are parameters which differ for each fragment-LCP pair. In
fact only 3 coordinates are needed to solve this equation, we then used
particular points from figure \ref{f:simon}.d to do so. The first one corresponds 
to the first point at which the difference function is equal to zero (at small relative velocity). 
The second point used is 
the local minimum
seen at small relative velocity (around 2.5 cm/ns) in the difference function (fig.\ref{f:simon}.d)
which corresponds to the minimum relative velocity allowed by the
Coulomb barrier. The third
one corresponds to the first point where the difference function is
equal to zero just after the second minimum, in this region the secondary
evaporation vanishes. 

In order to validate the method employed to estimate the background 
several tests have been made. We summarize the two most important 
tests that we already reported in ref.~\cite{I9-Mar97}: 

a) We compared the number of protons deduced by subtracting from the difference 
function (\ref{f:simon}.d) the real background and the background evaluated by the  
parametrisation of (Eq.1). We recover 91\% of the evaporated protons from Mg and 84\%
of evaporated protons from all prefragments.

b) The second check is related to the possible upper limit of the method. We performed 
SIMON simulations assuming higher excitation energies in the primary fragments. For 7.5 AMeV
excitation energies we recovered 81\% of evaporated protons. This result indicates that
the fraction of all evaporated protons recovered by this method is rather insensitive 
to the excitation energies of the primary fragments. 

Because the experimental
shape of the correlation function as well as the difference function
(Fig.\ref{f:data}) have the same behaviour as those in our simulation,
we applied the same method to the
experimental data to remove the background. From this simulation and 
method developed above we are able to isolate the LCP evaporated by
the primary fragments.

\subsection{Application to the data}

\begin{figure}
\includegraphics[width=\columnwidth]{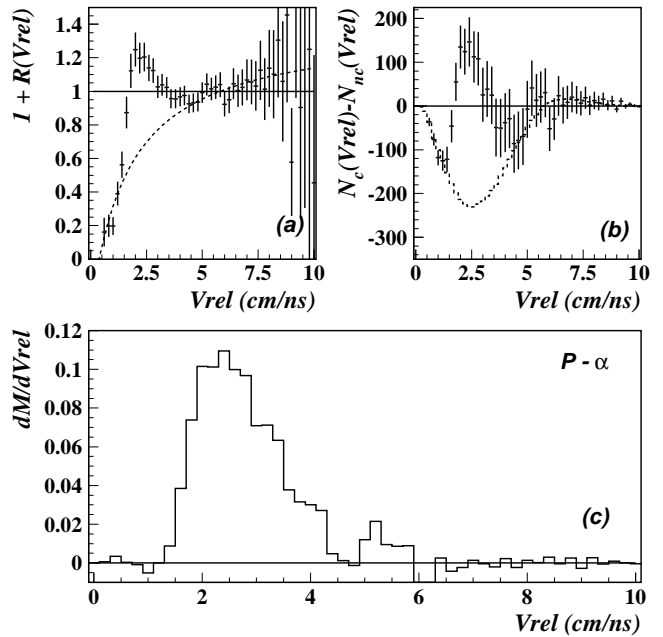}%
\caption{P-$\alpha$ correlation measured in central collisions of
Xe+Sn at 32 AMeV. (a) correlation function.
(b) difference function. (c) velocity spectrum of alphas in the
centre of mass of the Phosphorus fragment, obtained from the
subtraction of the difference function (data point in b)
and the background (dashed line in b).
\label{f:fond32}}
\end{figure}

Figure \ref{f:fond32} shows the experimental correlation function,
 the difference function
and the velocity distribution of $\alpha$ correlated to phosphorus fragments for the
central collisions of Xe + Sn at 32 AMeV. In the same figure are plotted the
corresponding background calculated with Eq.1 by using three points taken
from the experimental distributions as described in the above section.
Therefore the $\alpha$ velocity spectrum is deduced by subtracting the
background (the curve in fig.\ref{f:fond32} upper right panel ) from the 
difference 
function. This contribution represents the spectrum of $\alpha$ particles emitted by the
 parent of $P$ fragment. From the mean value 
of the distribution we 
can deduce the average kinetic energy of alpha. 
Its integral normalised 
to the total number of phosphorus nuclei provides the average multiplicity of $\alpha$
particles evaporated from parents of P fragments.

The uncertainties of the extracted quantities are mainly related to the
uncertainty of taking the three points which define the background. In practice
the first minimum in the difference function is easy to locate:
the corresponding 
error is small (see fig.\ref{f:fond32}.b).
The two other points are more difficult to extract, with the
possibility of significant uncertainties.
We then decided to take intervals around each point which are divided 
into a number of bins. Considering all possible combinations of one bin in the 
first interval and another in the second leads to a distribution of multiplicities.
This distribution has a narrow gaussian shape. We then consider the mean value
of this distribution
as the average multiplicity and its half-width as the error due to the method.    

\section{Experimental results}

\subsection{Average multiplicities and kinetic energies of the LCP
correlated to the fragments}\label{p:avg_mult}

\begin{figure}
\includegraphics[width=\columnwidth]{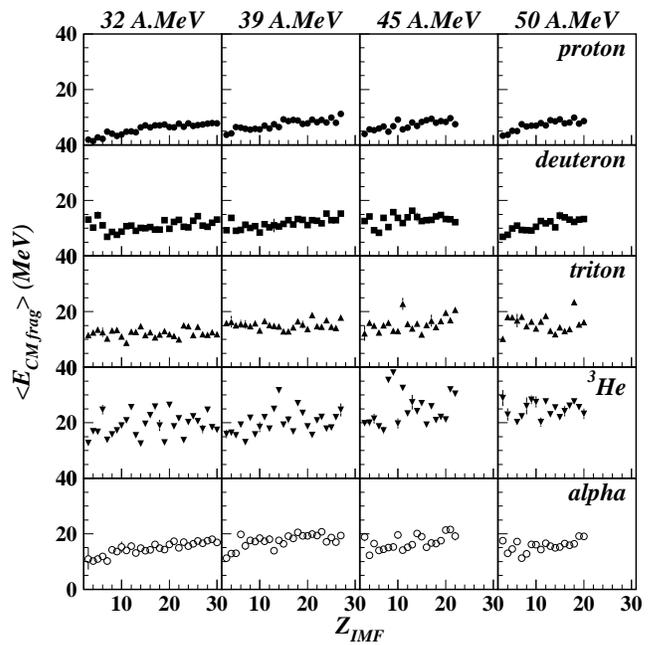}%
\caption{ Fragment centre of mass average kinetic energy of the secondary 
evaporated $p, d, t,^{3}He$ and $\alpha $  particles as a function of the 
atomic number of the fragments for central collisions of Xe+Sn at 32, 39, 
45 and 50 AMeV.
\label{f:e_pl}}
\end{figure}

We applied the method described above for all fragment-LCP pairs made by combining LCP 
isotopes ($p, d, t,^{3}He$ and $\alpha $)  and a range of fragments
emitted in central collisions between Xe and Sn at four 
incident energies, 32, 39, 45 and 50 AMeV. However due to a small cross section for
heavy fragment production which implies a low statistics 
(see figs. \ref{f:zdis} and \ref{f:zmax1})
 we performed these analyses for a limited
range of fragment charges depending on the beam energy. Thus the maximum
fragment charge we studied at 32 AMeV was 30, 27 at 39 AMeV, 22 at 45 AMeV and
20 for 50 AMeV.   
\begin{figure}
\includegraphics[width=\columnwidth]{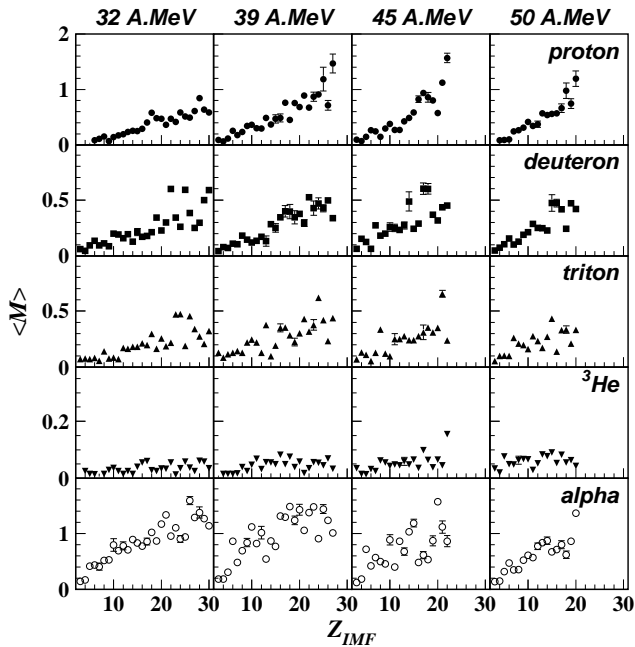}%
\caption{ Average secondary multiplicities per fragment of the evaporated
$p, d, t,^{3}He$ and $\alpha $  particles as a function of the atomic number of
the fragments for central collisions of Xe+Sn at 32, 39, 45 and 50 AMeV.  
The error bars correspond to the error due to the background extraction method.
\label{f:mult_pl}}
\end{figure}
The extracted average LCP multiplicities and their average kinetic energy  
 are given in figures \ref{f:mult_pl} and \ref{f:e_pl}
 as a function of the charge, $Z_{IMF}$, 
of the detected fragments and for the four bombarding energies. The average
multiplicities increase with the fragment size. The multiplicities are low and 
do not exceed a value of 1.5 which implies that the excitation energy of the
corresponding primary fragments is moderate. For a given light charged particle,
the multiplicity seems not to change with the beam energy.
From the spectra of the LCP evaporated from the parents of the detected
fragments we can extract the mean kinetic energy. This is shown in 
 figure \ref{f:e_pl}. It 
increases slightly with the charge of the fragment for the four incident
energies and in particular for proton and alpha particles. 
Notice that the kinetic energies of $^{3}He$ are
high compared to the values of the other particles. The observed effect
may be due to the higher identification threshold energy for $^{3}He$.

\subsection{Reconstruction of the size and excitation energy of the primary
fragments}
 
To reconstruct the charge of the primary fragments we used the LCP
multiplicities correlated to each fragment as described in the last paragraph.
Therefore the average charge of the primary fragment, $<Z_{pr}>$, is given
by the sum of the detected fragment and all evaporated LCP's charge weighted
by their corresponding multiplicities. $<Z_{pr}>$ is then given by the
relationship :
\begin{equation}
<Z_{pr}>=Z_{IMF}+\sum z_{i}<M_{i}>
\end{equation}
where $Z_{IMF}$ is the detected fragment charge, $z_{i}$ and $<M_{i}>$ are
the charge and the average multiplicity of the evaporated particle $%
i=p, d, t, ^{3}He$ and $\alpha $.

In order to reconstruct the mass of the primary fragments, a quantity needed to
deduce the excitation energy, we made two extreme assumptions : the first one
is that the primary fragments are produced in the valley of stability, the
second assumes that they are produced with the same N/Z ratio as the
composite initial system (N/Z conservation assumption). However, as mentioned above the INDRA 
detector does not
resolve the fragment isotopes, we therefore made an additional assumption which
supposes that the $Z$-identified detected fragments
have a mass corresponding to their valley of stability isotope.
In the framework of these assumptions we deduce from the primary fragment masses
the number of neutrons evaporated from the primary fragments.

\begin{figure}
\includegraphics[width=\columnwidth]{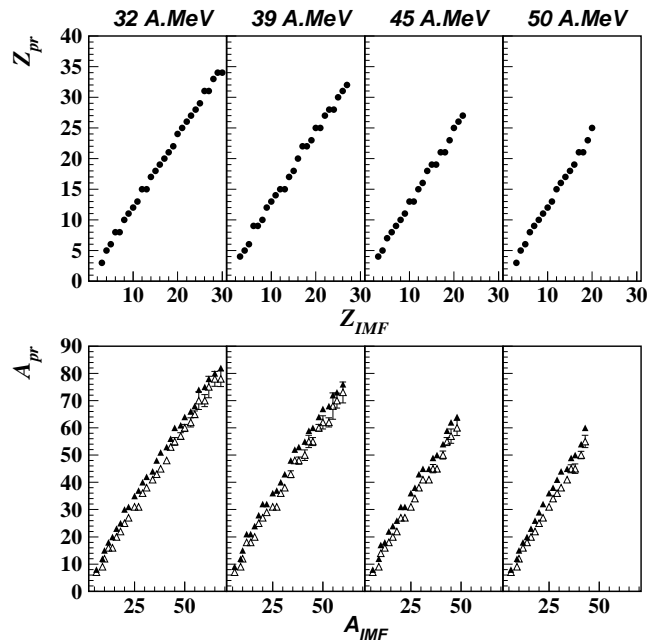}%
\caption{ The reconstructed charge and mass of the primary fragments as 
a function of the 
atomic number/mass of the detected fragments for central collisions of Xe+Sn at
32, 39, 45 and 50 AMeV. Two assumptions to reconstruct the masses are given: 
the open triangles correspond to the valley of stability case the black 
triangles represent the N/Z conservation hypothesis. (see text).
\label{f:zpr_apr}}
\end{figure}

\begin{figure}
\includegraphics[width=\columnwidth]{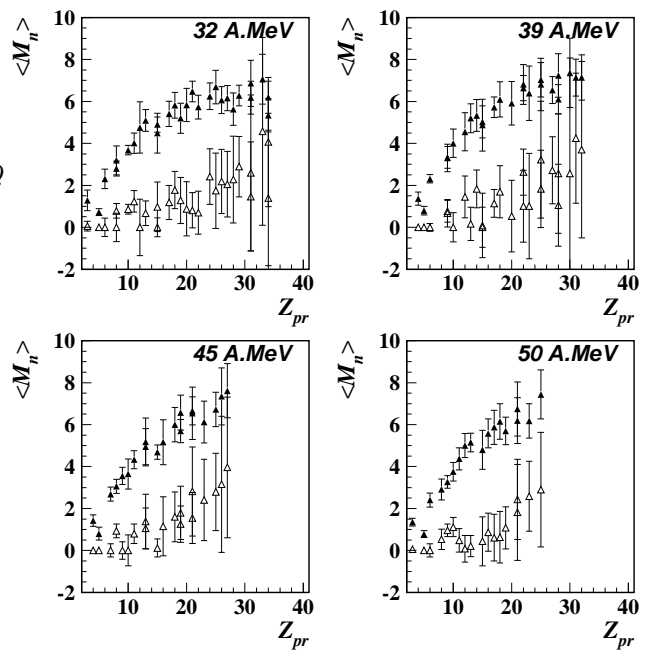}%
\caption{ The average neutron multiplicities evaporated from the primary
fragments deduced for central collisions of Xe+Sn at 32, 39, 
45 and 50 AMeV. Two assumptions to reconstruct the mass are given: the open
triangles correspond to the valley of stability case the black triangles
represent the N/Z conservation hypothesis (see text).
\label{f:mn}}
\end{figure}

Figures \ref{f:zpr_apr} and \ref{f:mn} show the result of this reconstruction
for the four incident energies. 
The values of the primary charge (fig.\ref{f:zpr_apr}, upper panel) obtained
vary from 1 to 5 charge units larger than the detected fragment. The mass
 of the
primary fragment depends on the assumption (fig.\ref{f:zpr_apr}, down panel).
The average neutron multiplicities are deduced from the mass conservation,
knowing the mass of the primary fragment, the detected fragment and the mass
of the secondary light charged particle contribution. Figure \ref{f:mn} shows for
the two assumptions the evolution of the number of neutrons for the four systems
as a function of the deduced primary fragment atomic number. Whatever the beam
energy, the multiplicity of neutrons reaches quite high values, up to 7 neutrons
for the  N/Z ratio conservation assumption. This is due to our assumption that 
detected fragments have 
their valley of stability mass. Clearly, when we also assume that the primary fragments are
produced in the valley of stability, the deduced neutron multiplicity cannot be very high. 
Conversely, imposing an N/Z of 1.39 for nuclei with Z = 3-30 means that primary fragments 
have large neutron excess compared to the (valley of stability) detected fragments.
     
At this stage, the calorimetric procedure can be applied to reconstruct the
average excitation energy of the primary fragments ($<E^*_{pr}>$). It is
given by the relationship :
\begin{equation}
<E^*_{pr}> = \sum<M_{LCP}><E_{LCP}>+<M_n><E_n> - Q
\end{equation}
where $<E_{LCP}>$ and $<E_n>$ are the average kinetic energies in the frame 
of the source (fragment) of the measured
evaporated LCP's and the deduced neutrons with the average multiplicity
$<M_n>$. The neutron kinetic energy $<E_n>$ is taken as the proton kinetic
energy minus the proton Coulomb barrier. $Q$ is the mass balance 
of the reaction.
\begin{figure}
\includegraphics[width=\columnwidth]{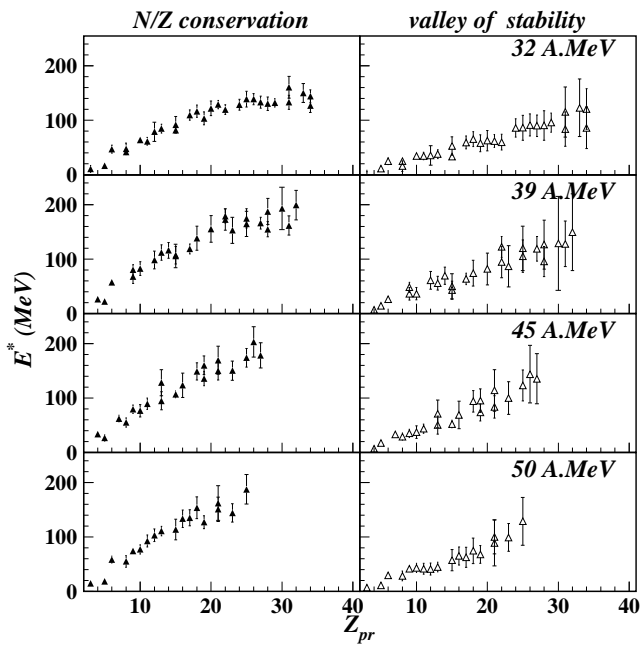}%
\caption{Average excitation energy of the primary fragments as a function of 
their atomic number for the central collisions of Xe+Sn at 32, 39, 
45 and 50 AMeV. Left panels: the primary fragments have the same N/Z as 
the combined system. Right panels: the fragments are produced in the valley 
of stability.
\label{f:excit}}
\end{figure}

\begin{figure}
\includegraphics[width=\columnwidth]{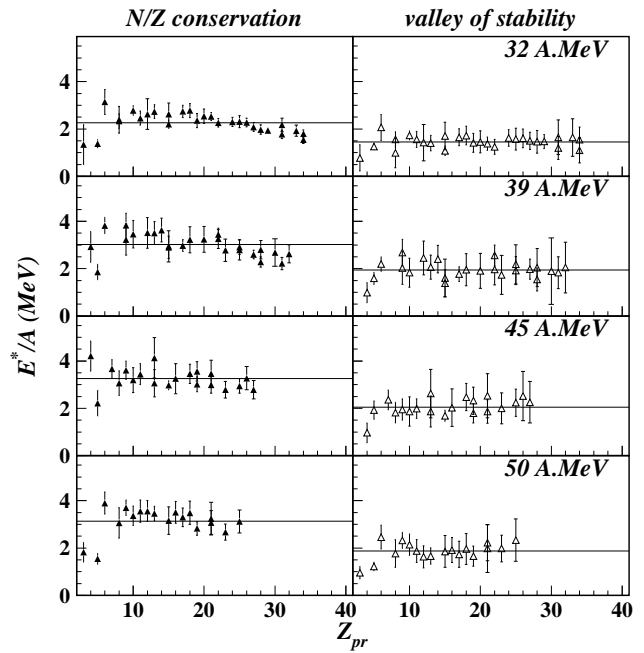}%
\caption{Average excitation energy per nucleon of the primary fragments 
as a function of 
their atomic number for the central collisions of Xe+Sn at 32, 39, 
45 and 50 AMeV. The horizontal lines represents the average value of the whole
set of the primary fragments. 
Left panels: the primary fragments have the same N/Z as 
the combined system. Right panels: the fragments are produced in the valley 
of stability. The vertical bars indicate errors due to the method. 
\label{f:esura}}
\end{figure}

Figure \ref{f:excit} shows the result of this procedure for the two
scenarios 
and at the four
bombarding energies. As expected from the deduced multiplicities (see
paragraph \ref{p:avg_mult}), the excitation energy increases with the size of the
 primary fragment for
all bombarding energies and for the two assumptions. However, for the 32 AMeV
system, $<E^*_{pr}>$ seems to saturate for the heavier fragments. We could 
wonder if this is due to limitations of the method. However, as we already
mentioned in paragraph \ref{p:sim_back}, simulations have been performed at 
much higher excitation energy into the primary fragments showing that we recover more than 80\%
of the evaporated protons. 

To decide between the scenarios for primary 
fragment mass, valley of stability or with the $N/Z$ conservation assumption, 
extensive statistical calculations have been performed using the 
GEMINI~\cite{cha88:Gemini} code, for the 50 AMeV system. 
In these calculations the input to the code was the experimental deduced 
primary charge, the fragment masses with the two assumptions and 
the associated excitation energies. The comparison to the experimental 
LCP multiplicities and kinetic energies suggests that the $N/Z$ conservation 
assumption 
is the most reasonable scenario. Details of these calculations are given in
ref.~\cite{I11-Mar98}. 

\begin{figure}
\includegraphics[width=.5\textwidth]{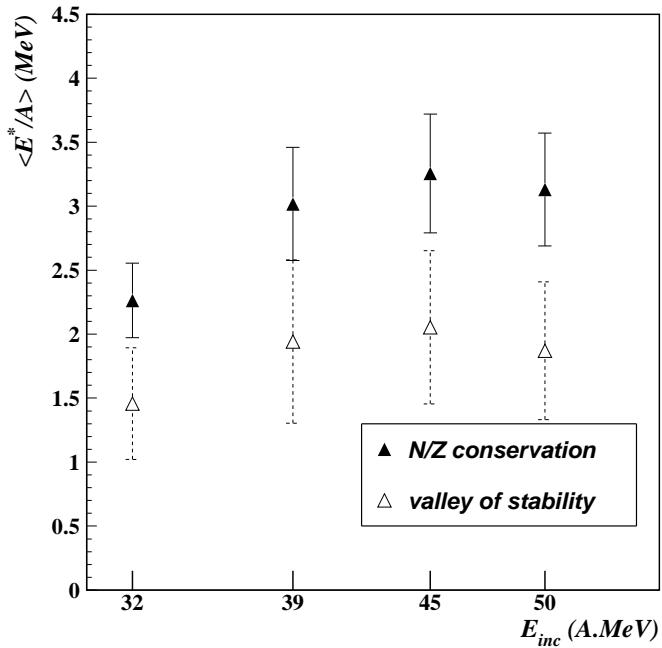}%
\caption{Average excitation energy per nucleon of the primary fragments 
as a function of bombarding energy. The black and open triangles correspond 
to the primary fragments having the same N/Z as the combined system, and 
produced in the valley of stability respectively. The error bars correspond to
the standard deviation from the mean values.
\label{f:ebeam_excit}}
\end{figure}

The linear trend of the $<E^*_{pr}>$ with the primary charge indicates
that the average excitation energy per nucleon, $<e^*_{pr}>$ in MeV/nucleon, 
is
constant whatever the size of the primary fragment. In figure \ref{f:esura} 
we verified the 
latter characteristic by plotting this variable. The horizontal lines in this
figure represent the average over the whole set of primary
fragments. Besides a few small 
charges the data points, within the error bars, lie on this straight horizontal 
line. 
Figure \ref{f:ebeam_excit} shows the evolution of
this value as a function of the bombarding energy. The vertical
bars are the standard deviations from the mean values. They are small and do
not exceed 1 AMeV, which supports the constancy of the value of $<e^*_{pr}>$.
For the N/Z conservation assumption the excitation energy per nucleon increases
from 2.2 AMeV at 32 AMeV and saturates at 3 AMeV beyond 39 AMeV. For the valley
of stability assumption, $<e^*_{pr}>$ saturates also but at a lower value.  
The constancy of the fragment excitation energy per nucleon, $<e^*_{pr}>$ 
for different fragment masses, seen in 
figure \ref{f:esura} has been interpreted in ~\cite{I11-Mar98} as meaning that, on the
average, thermodynamical equilibrium was achieved at the disassembly stage of
the system. Only one bombarding energy (50 AMeV) was available in the previous
work. On the other hand the saturation of $<e^*_{pr}>$ beyond 39 AMeV 
beam energy (fig.\ref{f:ebeam_excit}) may indicate that the fragments reach
their excitation energy limit 
(or limiting temperature)~\cite{Levit:T_limite,Koonin:T_limite}.

\subsection{Proportion of the evaporated light charged particles}

\begin{figure}
\includegraphics[width=\columnwidth]{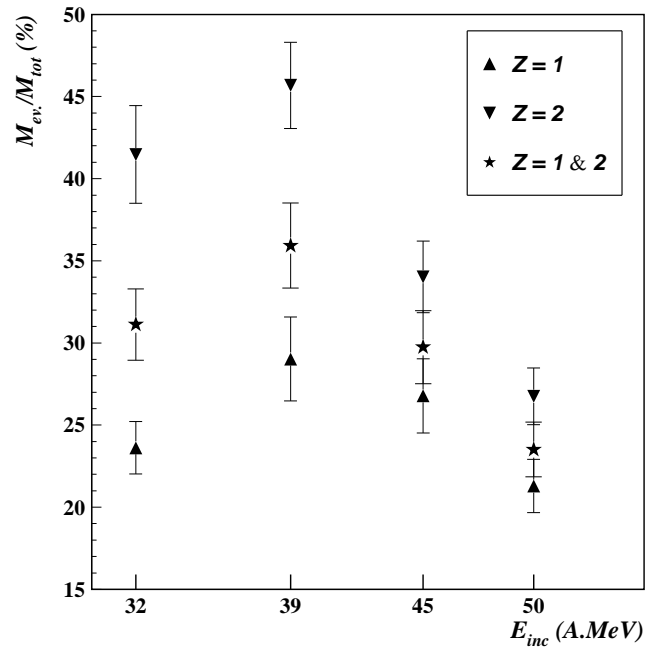}%
\caption{The ratios of the multiplicities per event of the secondary particles
evaporated by the primary fragments to the total emitted LCP vs the beam energy
for the Xe + Sn central collisions.
\label{f:pourc}}
\end{figure}

In paragraph \ref{p:avg_mult} we have extracted the average multiplicities of the 
secondary evaporated light charged particles for a given fragment. 
It is interesting to use this information in order to study the 
characteristics of the multifragmentation events. Indeed the LCP multiplicity
per event can be another pertinent observable. Table \ref{table1}
 shows the secondary LCP
multiplicities per event, the total LCP measured per event and the ratio of both
quantities, for the four beam energies. The secondary LCP multiplicities per
event are defined as the sum of the secondary evaporated LCP's per fragment, 
extracted by the method described
above, weighted by the measured fragment multiplicity per event, $M_{IMF}$.
These values are plotted in figure \ref{f:pourc} as a function of the beam 
energy. 

The fraction of helium isotopes evaporated in the decay of the primary
fragments is higher than for those of hydrogen. This difference is more pronounced
\begin{table*}
\begin{tabular}{|c||l||c|c|c|c|c|c|c|c|}
\hline 
\( E_{inc.} \)&
&
\( ^{1}H \)&
\( ^{2}H \)&
\( ^{3}H \)&
\( ^{3}He \)&
\( ^{4}He \)&
\( Z=1 \)&
\( Z=2 \)&
\( Z=1\&2 \)\\
\hline
\hline 
32&
\( M_{ev.} \) &
0.97&
0.83&
0.71&
0.12&
3.09&
2.51&
3.21&
5.72\\
\hline 
&
\( M_{tot} \)&
5.98&
2.85&
1.84&
0.38&
7.36&
10.67&
7.88&
18.55\\
\hline 
&
\( P_{ev.}\, (\%) \) &
\textbf{16.22}&
\textbf{29.12}&
\textbf{38.59}&
\textbf{31.58}&
\textbf{41.98}&
\textbf{23.52}&
\textbf{40.74}&
\textbf{30.84}\\
\hline
\hline 
39&
\( M_{ev.} \)&
1.73&
0.92&
1.1&
0.18&
4&
3.75&
4.18&
7.93\\
\hline 
&
\( M_{tot} \)&
7.16&
3.3&
2.45&
0.55&
8.6&
12.91&
9.15&
22.06\\
\hline 
&
 \( P_{ev.}\, (\%) \)&
\textbf{24.22}&
\textbf{27.95}&
\textbf{44.69}&
\textbf{32.36}&
\textbf{46.49}&
\textbf{29.06}&
\textbf{45.64}&
\textbf{35.94}\\
\hline
\hline 
45&
\( M_{ev.} \)&
1.68&
1.21&
1.01&
0.24&
3.2&
3.91&
3.44&
7.35\\
\hline 
&
\( M_{tot} \)&
7.82&
3.85&
2.93&
0.72&
9.39&
14.6&
10.11&
24.71\\
\hline 
&
 \( P_{ev.}\, (\%) \)&
\textbf{21.48}&
\textbf{31.51}&
\textbf{34.61}&
\textbf{33.89}&
\textbf{34.04}&
\textbf{26.76}&
\textbf{34.03}&
\textbf{29.73}\\
\hline
\hline 
50&
\( M_{ev.} \)&
1.42&
0.98&
1.01&
0.34&
2.6&
3.41&
2.94&
6.34\\
\hline 
&
\( M_{tot} \)&
8.37&
4.35&
3.3&
0.89&
10.1&
16.02&
10.99&
27.01\\
\hline 
&
\( P_{ev.}\, (\%) \)&
\textbf{16.99}&
\textbf{22.51}&
\textbf{30.45}&
\textbf{37.98}&
\textbf{25.71}&
\textbf{21.26}&
\textbf{26.71}&
\textbf{23.48}\\
\hline
\end{tabular}
\caption{\label{table1}Xe+Sn, central collisions : mean multiplicities of
evaporated particles per event. For each energy and particle,
\protect\( M_{ev.}\protect \) is the multiplicity of evaporated particles,
\protect\( M_{tot}\protect \) the total multiplicity,
\protect\( M_{ev.}/M_{tot}\protect \) the percentage of evaporated particles.}
\end{table*}
at lower beam energy. We observe also that the maximum proportion of evaporated particles 
does not exceed on the average 35\% of the total number of produced light charged
particles. The proportion of secondary particles increases between 32 and 
39 AMeV, which reflects the increasing of the excitation energy of the primary
fragments as it is seen in figure \ref{f:ebeam_excit}. 
Then this fraction decreases for higher incident energies, it reaches 23\% at 50
AMeV, while $<e^*_{pr}>$ saturates.

It should be noticed that the proportion of the secondary evaporated particles
given is a lower limit, because we did not consider the contribution that can
originate from the decay of unstable nuclei~\cite{nayak92:correl} such as $^8Be$, 
$^5Li$ etc. and the decay of short-lived excited sates.
 We finally have to stress that the results we obtained with the
 method described above are given on the average.

\section{Comparison with a statistical model}

An application of the experimental estimation of this secondary statistical component
is to constrain the statistical multifragmentation 
models~\cite{SMM:main-ref,SMM:botv87,MMMC:main-ref,Raduta:micro1,Raduta:micro2,
Parvan:stat,Pratt:stat}.
The comparison
of the extracted quantities with these models provides a crucial test of some of their
 basic assumptions.
Since in the MMMC~\cite{MMMC:main-ref} model the primary fragments undergo instantaneous
decay with neutron emission only, it can not be used for comparison with the data.

We have chosen to compare our data, with more details, to the SMM model using input source 
parameters very close to the ones already optimised in previous
works~\cite{bougault:Bormio97,Salou:PhD,len:PhD}. 
As shown there SMM provides a very good description of experimental fragment partitions.
In the present paper we aim to analyse the general behaviour of excitation energy of primary
fragments, therefore, for simplicity, the size of the initial source has
been fixed to be Z=83 and A=198 for the four incident energies. This
corresponds to N/Z = 1.39 which is the same ratio as the initial system.
This choice is justified by some dynamical calculations of source parameters
in this energy range \cite{Botvina:dynamical_nsurz,Tan:dynamical_nsurz}.
Although the $N/Z$ ratio of the SMM primary fragments increases slightly with increasing $Z$
of the fragments, it remains very close to the $N/Z$ ratio of the initial source \cite{Botvina:isotope}.
Therefore we will compare the results of these calculations to the extracted 
experimental results using the N/Z conservation hypothesis. The freeze-out volume has been
fixed to three times the normal volume. Finally for each incident energy 
we used the excitation energy of the initial source as a free parameter.
The thermal excitation energy values which reproduce best the charge 
distributions of the detected fragments are given in table \ref{excita smm}.

\subsection{Characteristics of the primary fragments}  

We have used a version of SMM where we have access to the freeze-out
configuration, i.e. to primary fragments' characteristics before secondary decay and Coulomb
propagation. 
This standard version is described in \cite{SMM:main-ref}.

\begin{figure}
\includegraphics[width=\columnwidth]{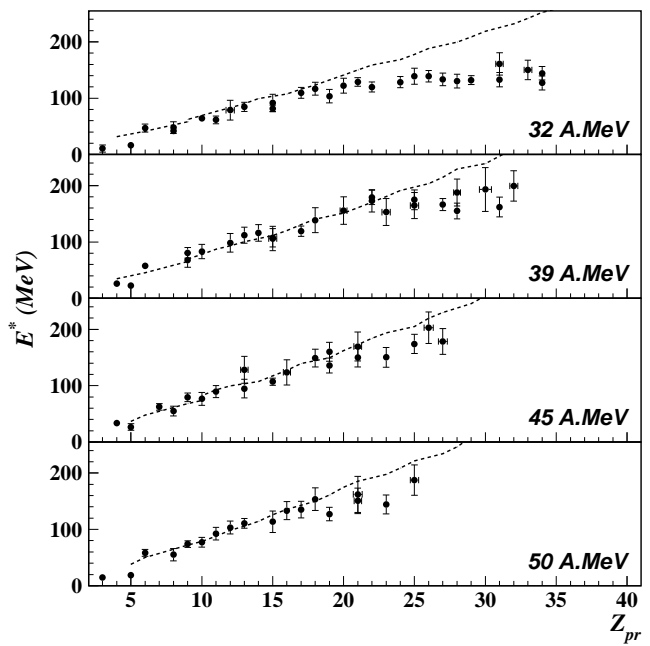}%
\caption{Average excitation energy of the primary fragments as a function of 
their atomic number for the central collisions of Xe+Sn at 32, 39, 
45 and 50 AMeV. The primary fragments are assumed to have same N/Z
ratio as the combined system. The symbols present the data and curves SMM
calculation.
\label{f:excit_smm}}
\end{figure}

The results of the SMM calculation, extracted directly from the freeze-out 
volume, are compared to the data in figure \ref{f:excit_smm}. 
The excitation energy of the primary fragments are
globally well reproduced for the four incident energies.
 
Small deviations are, however, observed for large primary fragment charges in 
particular for the 32 AMeV case. The experimental saturation of the excitation
energy is not reproduced.
Quantitative comparisons with the experimental excitation energy per nucleon 
averaged over the charge range of the measured fragments, are presented in 
table \ref{excita smm}.

\begin{table}[htb]
{\centering \begin{tabular}{|l|c|c|c|c|}
\hline 
Beam Energy (AMeV)&
32&
39&
45&
50\\
\hline
\hline
Thermal excitation energy&
5.&
6.&
6.5&
7.\\
\hline
\( \left\langle E^{\ast }/A\right\rangle _{exp.} \) (MeV)&
2.26&
3.02&
3.26&
3.13\\
\hline 
\( \left\langle E^{\ast }/A\right\rangle _{SMM} \) (MeV)&
2.97&
3.26&
3.39&
3.55\\
\hline
\end{tabular}\par}

\caption{\label{excita smm}
Thermal excitation energies in AMeV used in SMM simulations.
Experimental and calculated average excitation energies
of the primary fragments
produced in central collisions of Xe+Sn at 4 incident energies.}
\end{table}
The values of the calculated $<e^*_{pr}>$ show smooth increase with the beam energy
while the data seems to saturate at 3 AMeV above 39 A.MeV.

\subsection{Evaporated light charged particles}  

\begin{figure}
\includegraphics[width=\columnwidth]{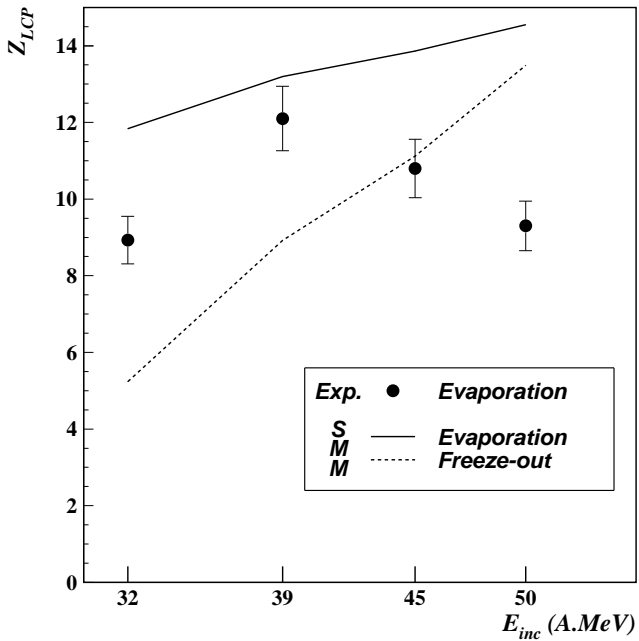}%
\caption{Total charge contributions of secondary evaporated particles 
and LCP produced at the freeze-out as a function of the beam energy.
The symbols represent the central collisions of Xe + Sn data,
evaporative part of SMM calculations is presented by 
 histogram and the freeze-out contribution by the 
dotted histogram 
\label{f:smm_lcp}}
\end{figure}

The contribution of the secondary evaporated LCP reflects the excitation 
energy of the primary fragments discussed in the previous paragraph. 
How do the small differences between the data and the calculation
for the excitation energy affect the predicted LCP multiplicities ?
We compare in figure \ref{f:smm_lcp} the charge contribution of total 
evaporated LCP resulting from SMM to the data.
      
The values extracted from the calculation are of the same order of magnitude
as the experimental ones. However the trend is not reproduced,  
SMM overestimates the evaporative contribution.

At 32 AMeV, the discrepancy can be due to 
the limited charge range considered experimentally.
Indeed, due to low statistics as we already mentioned, 
 we do not take into account evaporation from heavier fragments,
which are however more excited than lighter ones. To check this point
we extracted the total evaporated particles by using the 
correlation functions of reduced velocities instead of the relative
velocities ( $V_{red}=V_{rel}/\sqrt{Z1+Z2}$). This variable has the advantage  
to eliminate the charge dependence of the fragment-LCP 
relative velocity correlation functions~\cite{kim92:correl}. By doing this 
procedure, taking all fragments into account, the total charge of evaporated
  particles increases significantly to be in agreement with the calculated value
for the 32 AMeV case. 

For the 50 AMeV case, where the limitation in the charge range is more important
than for the other beam energies, the evaporated contribution changes very little 
and fails to increase the value of total evaporated charge from $Z_{LCP}$ = 9  
to $Z_{LCP}$ = 14 predicted by SMM (fig. \ref{f:smm_lcp}). 

The discrepancy is real, though, partly, it is caused by the thermal source size, which
should decrease with the beam energy. The SMM calculations do predict the decrease 
of the number of evaporated LCP, because of decreasing size and number of IMF at very large
excitation energies, in the ``falling'' part of the ``rise and fall'' of multifragmentation.
However, in the experiment
this effect is observed when the maximum of multifragmentation is not yet reached. 
In the calculations this behaviour takes place because the number of evaporated alpha
particles increases, contrary to the experimental result. This could be a consequence
of the secondary deexcitation prescription employed in SMM~\cite{SMM:botv87}. An other possible
reason would be an overexcitation of light primary IMF's predicted by SMM. The decay of these
IMF's contributes considerably into LCP production and their share increases with the 
thermal energy. 

The decrease of the experimental evaporated component $Z_{LCP}$ at high energy could be alternatively 
understood if we 
consider the increasing effect of the collision dynamics. The direct emission 
of LCP increases with the incident energy while the proportion of the thermal 
contribution decreases. This could be mocked up in the SMM calculations by
decreasing the thermal source size, but can in no case be predicted by SMM.

It is worth noting the contribution of light charged particles 
produced at freeze-out as predicted by SMM. Figure \ref{f:smm_lcp} 
shows that this contribution increases with the beam energy more rapidly than 
that of the evaporated particles. 
    
\section{Discussion of results}

In this work we have directly measured the saturation of the thermal excitation energy
deposited in fragments produced in central heavy-ion collisions between 32 and 50 AMeV,
by associating with each detected cold fragment the light charged particles evaporated by the
primary excited parent nucleus.
This saturation at excitation energies of around 3 AMeV observed in section \ref{s:extraction}
(see figure \ref{f:ebeam_excit}) is accompanied by
 a saturation of the number of evaporated LCP, that leads to 
 a decrease in the proportion of evaporated to all detected LCP,
with increasing incident energy (see figure \ref{f:pourc} or figure \ref{f:smm_lcp}). 

A similar saturation has been observed in an earlier work by Jiang et al.\cite{jiang89:T_limite}
using a completely different experimental method, based on the measurement of neutron
multiplicities.
The authors claimed the saturation of the thermal energy deposited in
hot nuclei formed in collisions of Ar + Au and Ar + Th in the energy range 27-77 AMeV.
Their claim was based on the observation of a saturation of the multiplicity of evaporated neutrons,
as well as that of the light charged particles detected in coincidence at backward angles, in central
collisions at increasing beam energies.
The neutron multiplicity saturates for the system Ar + Th around 30 AMeV at $<M_n> = 35$.
Let us note in passing that we estimate the neutron multiplicity per event evaporated by the system
Xe + Sn to be $<M_n> = 23$ at 39 AMeV. 

In \cite{jiang89:T_limite} the authors concluded that the observed saturation was
due to the increasing inefficiency of the reaction mechanism to deposit thermal energy in to the
hot nuclei it produced, rather than it being related to reaching the limits of excitation energy or
temperature that a nucleus may support. In discussing the results of the present work we must ask
ourselves the same question, but the situation is complicated by the fact that here we are dealing with
several heated nuclei per event which may themselves result from the break-up of some other
heavy, hot system. Here we will present some elements which may help to find an answer.

Although the excitation energies of primary fragments remain constant for incident energies
$\geq$39 AMeV, the detected fragment partitions continue to evolve, becoming steeper with
increasing bombarding energy (figure \ref{f:zdis})
while the average charge of the largest fragment varies from 25 at 32 AMeV to 15 at 50 AMeV
(figure \ref{f:zmax1}). This suggests that with increasing energy, above 39AMeV,
the average number and temperature of primary fragments produced in the reactions does not change, whereas
their average size decreases. Moreover, 
in order to conserve the total mass of the system the number
of light particles produced prior to secondary evaporation from fragments must also increase
with increasing energy.
If indeed the mechanism for thermal energy deposition saturates,
then the energy not used in forming and heating fragments has to be evacuated by some other
means, for example direct particle production. This would lead to such an increase in
non-evaporated particle multiplicity. Some energy may also be locked up as kinetic energy
of fragments due to some kind of collective motion, either isotropic (compression-expansion effects)
or anisotropic (incomplete stopping).

It should be recalled that the decrease of fragment excited state lifetimes with the excitation energy 
can limit the mean excitation values obtained in this paper. However, simulations we have performed
indicate that the effect of shorter lifetimes on the efficiency of the method is quite small for
primary fragment excitation energies up to 7.5 AMeV.

It is interesting to compare our results with a recent compilation of limiting temperatures  extracted
from different experimental measurements \cite{natow01:T_limite}.
It suggests that $T_{lim}$ decreases with increasing nuclear mass, in good agreement with
calculations \cite{zhang96:T_limite,zhang99:T_limite}. 
The primary fragments considered in the present work (figure \ref{f:zpr_apr})have, at the very
most, masses $A=80$, while most of them have masses in the region $A=10$--50.
The corresponding limiting temperature  from \cite{natow01:T_limite} is 
$T_{lim}=9$ MeV or $E^{*}/A=7.5$ AMeV.
As these values are much higher than the 3 AMeV maximum excitation energy we find in our
primary fragments, this would imply that the observed saturation is due to reaction mechanism and
not related to $T_{lim}$. 

However, in the same compilation limiting excitation energies $\leq 3$ AMeV are found for
the heaviest nuclei with masses in the $A=150$--200 range.
If we suppose that fragments are produced by the break-up of some heavy composite system
formed in the reaction (as in SMM calculations) then
it is possible that the observed saturation of primary fragment excitation
energies is due to the saturation of the excitation energy of the initial system, which attains its
(mass-dependent) $T_{lim}$.

\section{Conclusion}        

We have presented in this paper the experimental results of the 
intrinsic properties of the fragments produced in the
central collisions of Xe + Sn from 32 to 50 AMeV bombarding energy.   
Quantitative experimental determination of the size and excitation energy of the primary 
fragments produced at such 
collisions before their decay are given for the four beam energies. The comparison
of these extracted quantities with models provides a crucial test of some of their
basic assumptions.
 
The experimental methods used in this work are based on the relative velocity correlation 
functions between the detected fragment and light charged particles. Thus we have extracted
the average multiplicity of the evaporated particles and their average kinetic energies in the
centre of mass of the fragments. These two variables have been used in order to reconstruct the
average charge, mass and excitation energy of the primary fragments. 

Our results show that for a given beam energy, the excitation energy per nucleon is almost constant
over the whole studied range of fragment charge. The statistical multifragmentation model, SMM,
reproduces very well the internal excitation energy of the primary fragments. The average value 
of this quantity increase from 2.3 AMeV for a beam energy 32 AMeV  to saturate around 3 AMeV for 
39 AMeV and above.  

We also deduced the proportion of evaporated light charged particles per event, amounting to
30\% of the total measured LCP for the 32 AMeV reaction, increasing to 35\% at 39 AMeV
and decreasing down to 23\% for 50 AMeV. Therefore the majority of light charged particles
are not evaporated by excited primary fragments in these reactions. 
Neither the absolute values of this proportion nor its evolution are reproduced by SMM calculations
assuming a constant size for the multifragmenting system.
 
The two last results may indicate either i) that the system which disassembles into 
fragments is not able to sustain more than 3 AMeV thermal excitation energy and the excitation energy of the fragments 
reflects the temperature limit of that system, or/and ii) that the mechanism of dissipation of beam energy in to thermal
energy of hot nuclei saturates above $\sim 32$AMeV, and the kinetic energy in excess is evacuated via direct particle production
and collective motion of the fragments.
 
\begin{acknowledgments}
We thank A. Botvina for fruitful discussions and for performing the SMM
calculations.
We thank also the staff of the GANIL Accelerator facility for their support during
the experiment. This work was supported by Le Commissariat \`a l'Energie
Atomique, Le Centre National de la Recherche Scientifique, Le Minist\`ere de
l'Education Nationale, and le Conseil Regional de Basse Normandie. 
\end{acknowledgments}


\end{document}